\documentclass[conference]{IEEEtran}
\IEEEoverridecommandlockouts
\usepackage[bookmarks=false]{hyperref}
\usepackage{cite}
\usepackage{amsmath,amssymb,amsfonts}
\usepackage{algorithmic}
\usepackage{graphicx}
\usepackage{textcomp}
\usepackage{xcolor}
\usepackage{enumitem}

\usepackage{multirow}
\usepackage{array}
\usepackage{makecell}
\usepackage{booktabs}
\usepackage{placeins}

\usepackage{stfloats}

\def\BibTeX{{\rm B\kern-.05em{\sc i\kern-.025em b}\kern-.08em
    T\kern-.1667em\lower.7ex\hbox{E}\kern-.125emX}}
\begin{document}

\makeatletter
\newcommand{\linebreakand}{%
  \end{@IEEEauthorhalign}
  \hfill\mbox{}\par
  \mbox{}\hfill\begin{@IEEEauthorhalign}
}
\makeatother

\title{Forecasting Self-Similar User Traffic Demand Using Transformers in LEO Satellite Networks\\
}

\author{
\IEEEauthorblockN{ Yekta Demirci}
\IEEEauthorblockA{\textit{Poly-Grames Research Center, Dept. of Electrical Eng.} \\
\textit{Polytechnique Montréal,} Montréal, Canada \\
yekta.demirci@polymtl.ca}
\and
\IEEEauthorblockN{Guillaume Mantelet}
\IEEEauthorblockA{\textit{Satellite Systems, MDA} \\
Montréal, Canada \\
guillaume.mantelet@mda.space}
\and
\IEEEauthorblockN{Stéphane Martel}
\IEEEauthorblockA{\textit{Satellite Systems, MDA} \\
Montréal, Canada \\
stephane.martel@mda.space}
\linebreakand 
\IEEEauthorblockN{Jean-François Frigon}
\IEEEauthorblockA{\textit{Poly-Grames Research Center, Dept. of Electrical Eng.} \\
\textit{Polytechnique Montréal,} Montréal, Canada \\
j-f.frigon@polymtl.ca}
\and
\IEEEauthorblockN{Gunes Karabulut Kurt}
\IEEEauthorblockA{\textit{Poly-Grames Research Center, Dept. of Electrical Eng.} \\
\textit{Polytechnique Montréal,} Montréal, Canada \\
gunes.kurt@polymtl.ca}
}

\maketitle

\begin{abstract}
In this paper, we propose the use of a transformer-based model to address the need for forecasting user traffic
demand in the next generation Low Earth Orbit (LEO) satellite
networks. Considering a LEO satellite constellation, we present
the need to forecast the demand for the satellites in-orbit to utilize dynamic beam-hopping in high granularity. We adopt a traffic dataset with second-order self-similar characteristics. Given this traffic dataset, the Fractional Auto-regressive Integrated Moving Average (FARIMA) model is considered a benchmark forecasting solution. However, the constrained on-board processing capabilities of LEO satellites, combined with the need to fit a new model for each input sequence due to the nature of FARIMA, motivate the investigation of alternative solutions. As an alternative, a pretrained probabilistic time series model that utilizes transformers with a Prob-Sparse self-attention mechanism is considered. The considered solution is investigated under different time granularities with varying sequence and prediction lengths. Concluding this paper, we provide extensive simulation results where the transformer-based solution achieved up to six percent better forecasting accuracy on certain traffic conditions using mean squared error as the performance indicator.
\end{abstract}


\section{Introduction}
Low Earth Orbit (LEO) satellite constellations emerge as a potential solution to support ever-increasing broadband service demand across the globe. With the technological advancements in beamforming, advanced modulation techniques, reconfigurable phased array technologies, and electrically steerable antennas, the legacy bent-pipe satellites are evolving into extremely high-throughput satellites that embody regenerative payloads\cite{yahia2024evolution}. Embracing the Software Defined Networking (SDN) paradigms \cite{guo2022static} alongside the aforementioned advancements enables the satellites to be configured in-orbit and be responsive to the changing environmental conditions.   

In this work, we consider a LEO satellite constellation where each satellite with a regenerative payload provides broadband services to multiple earth-fixed cells. In each cell, there can be a different number of users with varying traffic demands on the forward link (downlink). Given this, the aggregated demand at each cell would be different. A snapshot of this system, where a single LEO satellite serving six fixed-earth cells can be seen on the left side of Fig.~\ref{fig:sats}. Considering this setting, different cells would experience varying, irregular traffic demands. Therefore, some cells may experience higher demand and become ``hot-spots" whereas some other cells may remain ``colder". In order to compromise this need, the Beam-Hopping (BH) technology emerges as a solution to utilize limited radio resources more flexibly compared to fixed beams. With the BH technology, hot-spot or higher priority cells can be illuminated for a longer duration of time with higher hopping rates.

There have already been some standardizations about how to schedule beams considering BH. DVB{-}S2X standards specify the notion of Beam Hopping Time Plans (BHTP) \cite{dvbs2x}. In this standardization, BHTP specifies how often each cell is illuminated, as well as the illumination duration of beams, also known as Dwell Time (DT). BHTP can be configured with fixed periodicity or can be traffic-driven as specified in ``Format 6". This format enables more flexible DT scheduling based on real-time traffic patterns instead of relying on static pre-scheduled plans. Yet, the traffic demand-based planning would require knowing the future user demand in advance as the plans are being generated.

\begin{figure*}[t]
\centerline{\includegraphics[width=0.8\linewidth]{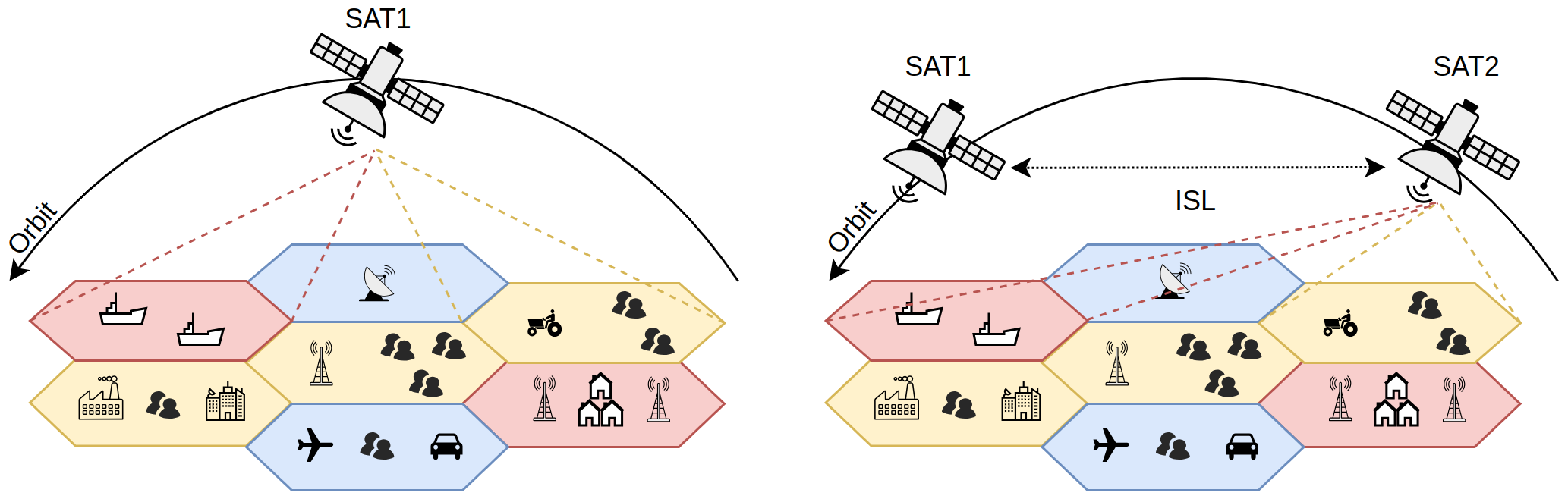}}
\caption{Two snapshots of the considered setting, in which beam hopping is used by LEO satellites to service fixed-earth cells with different user demands. On the left side, SAT1 serves six earth-fixed cells. On the right-hand side, SAT2 takes over serving the cells as SAT1 continues along its orbit.}

\label{fig:sats}
\end{figure*}

Considering the existing literature, there have been several works about how to allocate radio resources and create BH plans. In \cite{chen2024joint}, the authors proposed a BH scheduling framework considering power and time slot allocation. In \cite{lin2022multi}, a multi-satellite BH algorithm is proposed considering load balancing and intra-satellite interference. In \cite{wang2024joint}, the authors proposed an algorithm utilizing hybrid beamforming. All these works rely on knowing the user demand in advance and assume the future user demand is pre-known. Yet this remains as a literature gap. Additionally, the users at a given cell generate packets in millisecond time granularity, whereas BH plans are generated in seconds to minutes granularity. As a result, when user demand increases unexpectedly, low-granularity BH plans may fail to adapt promptly, leading to varying jitter and even packet loss due to limited on-board buffering. This highlights the need for accurate, high-granularity forecasting of user demand for upcoming time slots.

 Besides the aforementioned one, there remains another research gap. Considering the LEO satellite are non-geostationary, a satellite can provide service to a cell for 5-10 minutes before it hovers over and starts serving the next cells on its orbital trajectory. In order to provide seamless service to the users in the cell, another LEO satellite from the constellation emerges and continues the service. A snapshot of this process can be seen on the right side of Fig.~\ref{fig:sats}. During this period, the departing satellite (SAT1) may transmit metadata—such as user demand information collected during its service—to the incoming satellite (SAT2) via the Inter-Satellite Link (ISL). Yet, this process would take some time. Therefore, there is a need to forecast multiple steps ahead in advance during this process so that BHTP can be generated without any disruption. Therefore, there is a need to model the user traffic and forecast the demand before creating any BHTP.

There have been some works to model and simulate the user traffic in LEO satellite constellations. In \cite{al2020traffic}, the authors proposed a traffic simulator for the satellite systems, considering three main types of users: i) land, ii) maritime, and iii) aeronautical. However, their time granularity is in hours, and the model is not feasible to analyze user demand in second-minute intervals, in other words, for higher granularity. In \cite{guo2021gateway}, the authors used a gravity model to estimate user traffic between different areas; however, their concern is over long time intervals to find gateway placement locations, which is not useful for demand forecasting in high granularity. In \cite{bie2019combined}, the authors considered a self-similar traffic model, yet their work does not provide a solution for multiple-step forecasting nor run-time complexity analysis to investigate the feasibility of the proposed solution in-orbit. Targeting the aforementioned gaps in the literature, our contributions are listed as follows:\\
\begin{enumerate}[label=C-\arabic*]
\item We identify the need for forecasting traffic demand in-orbit for time intervals with high granularity to realize the next generation high-throughput low latency LEO satellites.
\item We propose the use of a transformer-based forecasting model with Prob-Sparse attention, achieving $\mathcal{O}(L \log L)$ run-time and memory complexity as well as utilizing a Graphics Processing Unit (GPU) that can be feasible for in-orbit forecasting.
\item We provide comprehensive test results with varying time granularities and different prediction horizons, demonstrating the superior forecasting accuracy of the considered model based on the Mean Squared Error (MSE) metric, compared to Autoregressive Integrated Moving Average (ARIMA) and Fractional ARIMA (FARIMA).
\end{enumerate}

The following parts of the paper are organized as follows. Details of the second-order self-similar traffic model and the benchmark forecasting solutions ARIMA and FARIMA are given in Section \ref{background}. The proposed transformer-based solution is given in \ref{transformer} with a time-complexity analysis, and the simulation results are presented in \ref{simulations}. Then the paper is concluded with Section \ref{conc}

\section{Background} \label{background}
To characterize user demand within fixed Earth cells, we begin by examining a traffic model. However, to the best of our knowledge, there is no publicly available LEO satellite network dataset to create such a model. This encouraged us to investigate existing traffic models that have some underlying similarities with the emerging satellite networks. Considering the aggregated nature of the demand at each cell and the LEO satellites providing broadband service mostly using Internet Protocol (IP), we adopt a self-similar traffic model \cite{leland2002self}.
\subsection{Second Order Self Similar Traffic Model} \label{traffic}
Leland et al. \cite{leland2002self} showed that Ethernet LAN traffic shows a self-similar characteristic. The degree of self-similarity can be measured using the Hurst parameter, which describes the burstiness of a given dataset. Following their notion, let $X=(X_t: t = 0,1,2,...)$ be a covariance stationary stochastic process with mean $\mu$, variance $\sigma^2$ and an autocorrelation function $r(k), k\geq0$. For $0<\beta<1$, $r(k)$ is assumed to have the form:
\begin{equation}
r(k) \sim k^{-\beta} L(t) \text{ , } k \rightarrow \infty.
\label{eq:first}
\end{equation}
$L(t)$ is a slowly varying function at infinity for all $x>0$
\begin{equation}
    \text{lim}_{t\rightarrow \infty} \frac{L(tx)}{L(t)} = 1.
\end{equation}
Let $X^{(m)}=(X_k^{(m)}:k=1,2,3,...)$ denote new covariance stationary time series obtained averaging $X$ over non-overlapping blocks of size $m$. More formally:
\begin{equation}
    X_k^{(m)} = \frac{1}{m}(X_{km-m+1}+...+X_{km}), \text{ }k\geq1.
\end{equation}
Then the process $X$ is called (exactly) second-order self-similar if it holds both \eqref{eq:third}
 and \eqref{eq:fourth} or called (asymptotically) second-order self-similar if it holds \eqref{eq:fifth} for all $k$ large enough:
 \begin{equation}
\text{var}(X^{m})=\sigma^2m^{-\beta} \text{ , } \forall m
 \label{eq:third}
 \end{equation}
 \begin{equation}
     r^{(m)}(k)=r(k) \text{ , } k\geq0
\label{eq:fourth}
 \end{equation}
\begin{equation}
    r^{(m)}(k) \rightarrow r(k) \text{ , } m \rightarrow \infty. 
\label{eq:fifth}
\end{equation}
Given a process is second-order self-similar, its self-similarity can be parametrized with the Hurst parameter, $H$, as given $H=1-\beta/2$.
In their work\cite{leland2002self}, Leland et al. investigated some Ethernet traffic data that was sampled over a 20-40 hour interval where a Hurst parameter was found for the entire interval. Based on the International Telecommunication Union (ITU) report \cite{iturm2370}, which provides traffic estimates for 2020–2030, we observe that application traffic demand in the North American Region (NAR) exhibits non-stationary behavior over 24 hours. To capture this, we adopt a self-similar traffic model with a time-varying Hurst parameter that reflects changes across different service durations. In other words, the Hurst parameter would vary for different service time intervals. Service duration is considered the time being a satellite providing broadband service to a given cell (\textit{e.g.} 10 minutes). We identify three time intervals where the demand can be considered as low, medium, and high, as represented in Fig.~\ref{fig:demand} with blue, yellow, and red, respectively.

Given this traffic model, we have generated synthetic traffic data using \textit{ON} and \textit{OFF} packet trains \cite{taqqu1997proof}. Suppose there are \textit{M} independent and identically distributed (i.i.d.) sources where each generates binary time series $\{W(t), t\geq0\}$, $W(t)=1$ being there is a packet at time t and $W(t)=0$ being there is not. The superposition of such \textit{M} sources rescaled by a factor $T$ gives the total packet count at $t$ as:

\begin{equation}
    W^{*}_{M}(Tt) = \int_{0}^{Tt}( \sum_{m=1}^M W^{(m)}(u) )du.
\label{eq:superpose}
\end{equation}

\begin{figure}[t]
\centerline{\includegraphics[width=0.8\linewidth]{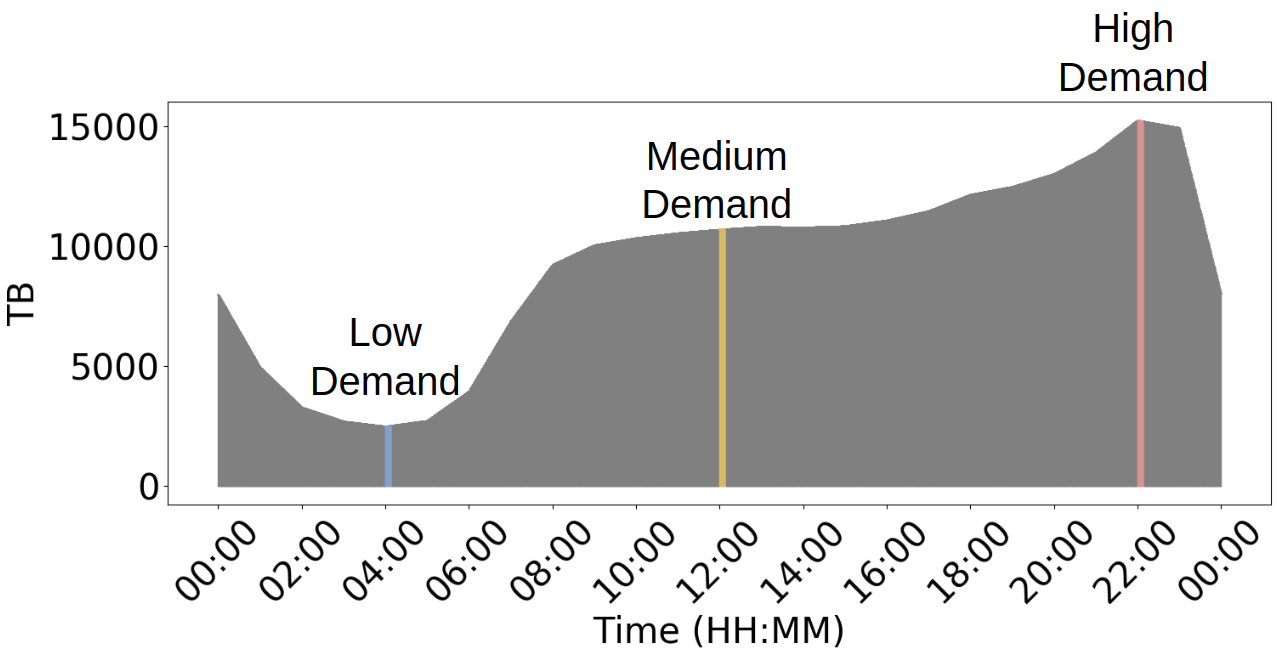}}
\caption{Daily application traffic profile for North American Region based on the data from \cite{iturm2370}}
\label{fig:demand}
\end{figure}
Assume $f_i(x)$ is the probability density function for \textit{ON/OFF} periods for $i=1$ representing \textit{ON} and $i=2$ representing \textit{OFF}. Let $F_i(x) = \int_0^x f_i(u)du$ be the cumulative distribution function and its complementary $F_{ic}=1-F_i(x)$.
The mean is $\mu_i=\int_0^\infty xf_i(x)dx$ and the variance is $ \sigma_i^2=\int_0^\infty(x-\mu_i)^2f_i(x)dx$. Assume as $x\rightarrow \infty$, either \eqref{eq:cond1} or $\sigma^2_i<\infty$  holds for $i \in \{1,2\}$ where $l_i>0$ is a constant and $L_i>0$ is a slowly varying function at infinity being $\text{lim}_{x\rightarrow\infty}L_i(tx)/L_i(x)=1$ for any $t>0$:
\begin{equation}
    F_{ic}(x)\sim l_ix^{-a_i}L_i(x) \text{ , } 1<a_i<2.
\label{eq:cond1}
\end{equation}
$a$ is the shape parameter of the respective distribution. For large \textit{M} and \textit{T} values, the stochastic process given in  \eqref{eq:superpose} can be adequately normalized to $\{\sigma_{lim}B_H(t),  t\geq0\}$ \cite{taqqu1997proof}. The normalization comes from the limits in \eqref{eq:limits} where $\mathcal{L} \lim$ means convergence in the sense of the finite-dimensional distributions. The interested readers can refer to the cited paper for the detailed proof of:
\begin{equation}
    \mathcal{L} \lim_{T \rightarrow \infty} \mathcal{L} \lim_{M \rightarrow \infty} \frac{\left(W^*_M(Tt)-TM\frac{\mu_1}{\mu_2+\mu_2}t\right)}{T^HL^{1/2}(T)M^{1/2}}=\sigma_{lim}B_H(t).
\label{eq:limits}
\end{equation}

\begin{figure*}[t]
\centerline{\includegraphics[width=0.8\textwidth]{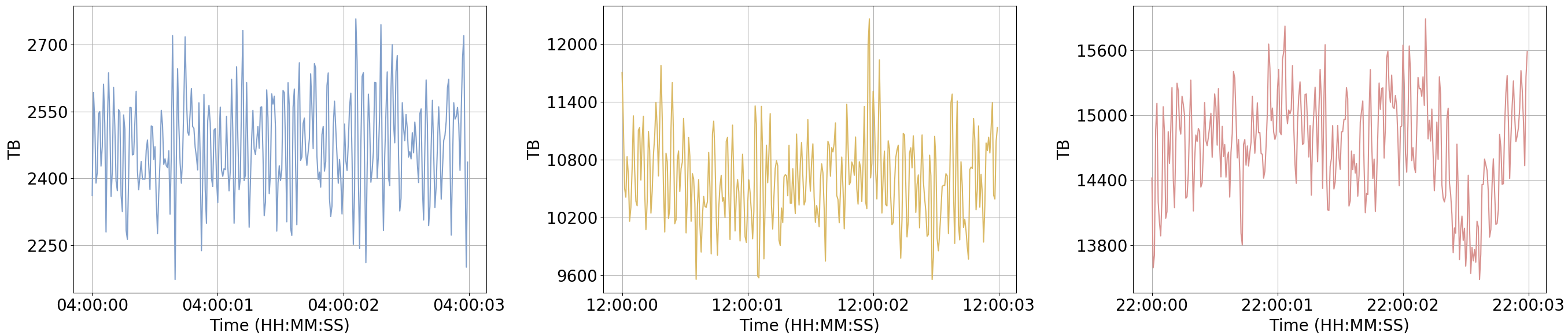}}
\caption{A snapshot of some synthetically generated data for the low, the medium, and the high demand intervals considered in Fig. \ref{fig:demand}.}
\label{fig:demandGranular}
\end{figure*}

$\sigma_{lim}$ is a finite positive constant and $B_H$ is fractional Brownian Motion, which is a Gaussian Process with stationary increments that is self-similar. $B_H$ has the following covariance function where $H$ is the Hurst parameter:
\begin{equation}
    \text{cov}(B_H(s),B_H(t))=\frac{1}{2}\left\{s^{2H}+t^{2H}-|s-t|^{2H}\right\}.
\label{eq:cov}
\end{equation}

The relationship between the shape and Hurst parameters is given as $H=(3-a_{min})/2$. One important thing to note is that the source with the highest $H$, or equivalently with the smallest $a$, ultimately dominates as $T\rightarrow\infty$ as pointed out in \cite{taqqu1997proof}. In other words, only the smallest $a$ parameter among the $M$ sources concerns us. Using \textit{ON/OFF} packet trains and following the values given \cite{iturm2370}, it is possible to create synthetic data for the considered service time intervals with high granularity (\textit{e.g.} 10 milliseconds), as it can be seen in Fig. \ref{fig:demandGranular}.

\subsection{ARIMA}\label{arima}
ARIMA is a linear model for time series analysis. An ARIMA$(p,d,q)$ process can be defined by \eqref{eq:arima} following Box and Jenkins' notation \cite{beran2013long} where $(1-B)^d$ represents the differencing operator, $d \in\{1,2,3...\}$
\begin{equation}
    (1-B)^dZ_t=Y_t \text{, } t\geq1.
\label{eq:arima}
\end{equation}
$Y_t$ can be modeled with Autoregressive Moving-Average (ARMA). An ARMA$(p,q)$ process is given by \eqref{eq:arma} where $\varepsilon_t$ is assumed to be i.i.d. with zero mean and finite variance $\sigma_{\varepsilon}^2$. $\phi(k)=1-\sum_{j=1}^p\phi_jk^j$ and $\psi(k)=\sum_{j=0}^q\psi_jk^j$ are polynomials with no common roots and all roots being outside the unit circle \cite{beran2013long}:
\begin{equation}
    \phi(B)Y_t=\psi(B)\varepsilon_t.
\label{eq:arma}
\end{equation}
\subsection{FARIMA}\label{farima}
When $d$ in \eqref{eq:arima} is allowed to be non-integer values, the ARIMA process can be extended to a FARIMA process. Let $d \in (- \frac{1}{2},\frac{1}{2})$, then the fractional differencing can be computed with the following series expansions \cite{beran2013long}:
\begin{equation}
    A^{-1}(k)=(1-k)^{d} = \sum_{j=0}^\infty a_jk^j.
\label{eq:infSeries}
\end{equation}
For $|k|\leq1$, $k\neq1$ and $\Gamma(\cdot)$ being Gamma function:
\begin{equation}
    a_j=\frac{\Gamma(d+1)}{\Gamma(j+1)\Gamma(d-j+1)}(-1)^j.
\end{equation}

FARIMA processes are asymptotically second-order self-similar with self-similarity parameter $d+1/2$, given $0<d<1/2$ \cite{leland2002self}. Therefore, we consider a FARIMA-based forecasting as the benchmark solution in our work.
\section{Transformer based forecasting solution} \label{transformer}
Neural networks with transformer architectures emerged as strong pattern recognizers \cite{vaswani2017attention}. Initially, they were proposed for text sequence-to-sequence (seq2seq) problems. Yet, their strong pattern recognition capabilities have since been adopted in many different domains, from computer vision to time series analysis \cite{fournier2023practical}. The attention mechanism allows the long dependency to be captured by relating any two positions of a given input sequence. 

For each input sequence, there exists a tuple of Query, Key, and Value as shown by $Q, K, V$ respectively. The relationship between the input sequence and the $Q, K, V$ depends on the dimension of the model. Given $softmax(\cdot)$, a normalization function to obtain probability distributions, the attention mechanism is given as:
\begin{equation}
    \text{Attention}(Q,K,V) = softmax\left(\frac{QK^{\top}}{\sqrt{d_k}}\right)V,
\label{eq:attention}
\end{equation}
where $d_k$ constant represents the dimension of $Q$ and $K$ \cite{vaswani2017attention}. Considering this, the $QK^{\top}$ operation has a quadratic run-time and memory complexity in vanilla transformers.

Zhou et al. \cite{zhou2021informer} addressed this shortcoming by proposing a Prob-Sparse attention mechanism achieving $\mathcal{O}(L \log L)$ for long sequence time-series forecasting. During the attention mechanism, instead of using all $L$, the proposed architecture uses only the top-$u$ queries where $u=c \ln{L_Q}$ and $c$ is a constant. They employ a max-mean criterion to select the top-$u$ queries, ultimately achieving $\mathcal{O}(L \log L)$ run-time and space complexity. We consider this architecture, Informer, as a transformer-based solution.

\subsection{Complexity analysis of the considered forecasting solutions}
Run-time complexity of the considered solutions are as follows:
\subsubsection{ARIMA/FARIMA}
To obtain a FARIMA$(p,d,q)$ model for an input sequence of length $L$, the four-step procedure in \cite{montanari1997fractionally} can be used:
\begin{enumerate}
\item Estimate the preliminary value of $d$ using rescaled-range analysis, which has a run-time complexity of $\mathcal{O}(L \log L)$.
\item Apply fractional differencing by truncating the infinite series in \eqref{eq:infSeries} to obtain an ARMA model. A naive approach has $\mathcal{O}(L^2)$ complexity, but this can be reduced to $\mathcal{O}(L \log L)$ using FFT-based convolution \cite{veenstra2013persistence}.

\item \label{step} Determine $p$ and $q$. Although Partial Autocorrelation Function  (PACF) based heuristics can be used to to determine $p$ and $q$ \cite{veenstra2013persistence}, they are infeasible for in-orbit use. Alternatively, model fitting over candidate values $p^\dagger = q^\dagger$ can be done, but this results in $\mathcal{O}({p^\dagger}^2)$ complexity, which would be impractical for in-orbit deployment.\label{arfar}

\item Estimate ARMA parameters via Maximum Likelihood Estimation using the Durbin-Levinson algorithm $\mathcal{O}(L^2)$. For longer sequences, Whittle Likelihood estimation can reduce this to $\mathcal{O}(L \log L)$ \cite{veenstra2013persistence}.

\end{enumerate}

In summary, forecasting a sequence of length $L$ using a FARIMA model has a run-time complexity of $\mathcal{O}(L \log L)$ for a given $p$ and $q$. However, selecting optimal $p$ and $q$ introduces additional overhead (step~\ref{step}), making it unsuitable for in-orbit implementation. ARIMA follows a similar procedure but omits fractional differencing.

\subsubsection{Informer} Informer predicts a sequence at one forward operation rather than step-by-step. For a given Query and Key pair with each having a dimension of $L$, the run-time complexity of forecasting is $\mathcal{O}(L \log L)$ \cite{zhou2021informer}. It is worth noting that the dimensions of the Query and Key matrices depend not only on the input sequence length but also on the model dimension.

\section{Simulation Results} \label{simulations}

To account for the varying traffic demand throughout the day, as shown in Figure~\ref{fig:demand}, we adopt shape and Hurst parameter pairs $(a, H)$ of (1.04, 0.98), (1.6, 0.7), and (1.9, 0.55) to represent high, medium, and low demand scenarios, respectively, inspired by \cite{leland2002self}. These $(a)$ values approximately capture the two ends of long-range and short-range dependence, considering \eqref{eq:cond1}. Although the low-demand scenario in Figure~\ref{fig:demand} may exhibit a stronger long-range dependency than what $a = 1.9$ reflects, using these extreme values allows us to conduct a more comprehensive analysis. Based on the assumed traffic demand, we set the number of terminals $M$ within a single cell to 750, 500, and 250 for high, medium, and low demand cases, respectively. Each terminal contributes 1~Mbps of traffic while in the \textit{ON} state.

\renewcommand{\arraystretch}{1.5} 
\setlength{\tabcolsep}{4pt} 
\newcolumntype{M}[1]{>{\centering\arraybackslash}m{#1}}

\begin{table*}[!b]
\centering
\caption{Table for the high demand traffic; $M=750$, $a=1.04$, $H=0.98$}
\resizebox{\textwidth}{!}{%
\begin{tabular}{!{\vrule width 1.1pt}  M{0.6cm}|M{0.6cm}|M{1cm} !{\vrule width 1.1pt}  *{3}{c|} c !{\vrule width 1.1pt} *{3}{c|} c!{\vrule width 1.1pt} *{3}{c|} c!{\vrule width 1.1pt}   }

\Xhline{1.1pt}
\multicolumn{3}{ !{\vrule width 1.1pt}  c !{\vrule width 1.1pt} }{ Granularity } & \multicolumn{4}{c  !{\vrule width 1.1pt}}{\textbf{10ms}} & \multicolumn{4}{c  !{\vrule width 1.1pt}}{\textbf{100ms}} & \multicolumn{4}{c !{\vrule width 1.1pt} }{\textbf{1000ms (1s)}} \\
\hline
\multicolumn{3}{ !{\vrule width 1.1pt}  c !{\vrule width 1.1pt} }{ Sequence Length } & 64 & 128 & 256 & 512 & 64 & 128 & 256 & 512 & 64 & 128 & 256 & 512 \\
\Xcline{1-4}{0.4pt}
\Xcline{4-15}{1.1pt}

&  & INF & \textbf{0.3434} & \textbf{0.3413} & \textbf{0.3401} & \textbf{0.3404} & \textbf{0.1892} & \textbf{0.1881} & \textbf{0.1879} & \textbf{0.1876}  & \textbf{0.2422} & \textbf{0.2401} & \textbf{0.2396} & 0.2410 \\ \noalign{\vskip -6pt}
& 1 & FAR & 0.3688 & 0.3513 & 0.3446 & 0.3417 & 0.2006 & 0.1923 & 0.1897 & 0.1877 & 0.2546 & 0.2458 & 0.2403 & \textbf{0.2383}  \\ \noalign{\vskip -6pt}
& & AR & 0.4154 & 0.4184 & 0.4247 & 0.4298  & 0.1979 & 0.1944 & 0.1938 & 0.1951 & 0.2515 & 0.2485 & 0.2461 & 0.2463  \\ \noalign{\vskip -1pt}
\Xcline{2-3}{0.4pt}
\Xcline{4-15}{0.4pt}

\multirow{5}{*}{\rotatebox{90}{Prediction Length}}&
& INF & \textbf{0.6090} & \textbf{0.6083} & 0.6185 & 0.6235 &   \textbf{0.3724} & \textbf{0.3725} & \textbf{0.3709}  & 0.3807 & \textbf{0.4717} & \textbf{0.4694} & 0.4822 & 0.4675  \\ \noalign{\vskip -6pt}
 & 12 & FAR & 0.6318 & 0.6126 & \textbf{0.6044} & \textbf{0.5999} & 0.4056 & 0.3882 & 0.3827 & \textbf{0.3700}  & 0.5111 & 0.4827 & \textbf{0.4665} & \textbf{0.4557} \\ \noalign{\vskip -6pt}
& & AR & 0.7027 & 0.7340 & 0.7750 & 0.8098 & 0.3994 & 0.4047 & 0.4185 & 0.4397 & 0.5007 & 0.5070 & 0.5084 & 0.5194 \\ \noalign{\vskip -1pt}
\Xcline{2-3}{0.4pt}
\Xcline{4-15}{0.4pt}

&  & INF & \textbf{0.6683} & \textbf{0.6713} & 0.6811 & 0.6727  & \textbf{0.4338} & \textbf{0.4292} & \textbf{0.4349} & 0.4514  & \textbf{0.5441} & \textbf{0.5522} & 0.5679 & 0.5562 \\ \noalign{\vskip -6pt}
& 24 & FAR & 0.6903 & 0.6759 & \textbf{0.6678} & \textbf{0.6635} & 0.4617 & 0.4474 & 0.4462 & \textbf{0.4287} & 0.5794 & 0.5544 & \textbf{0.5368} & \textbf{0.5249} \\ \noalign{\vskip -6pt}
& & AR & 0.7410 & 0.7674 & 0.8065 & 0.8421 & 0.4597 & 0.4653 & 0.4914 & 0.5315 & 0.5767 & 0.5814 & 0.5869 & 0.6068 \\ \noalign{\vskip -1pt}
\Xcline{2-3}{0.4pt}
\Xcline{4-15}{0.4pt}

&  & INF & \textbf{0.7270} & \textbf{0.7306} & 0.7409 & 0.7491 & \textbf{0.4926} & \textbf{0.5015} & \textbf{0.4920} & 0.5019 & \textbf{0.6394} & 0.6299 & 0.6554 & 0.6086 \\ \noalign{\vskip -6pt}
& 48 & FAR & 0.7582 & 0.7454 & \textbf{0.7332} & \textbf{0.7303} & 0.5219 & 0.5042 & 0.5128 & \textbf{0.4894} & 0.6542 & \textbf{0.6252} & \textbf{0.6078} & \textbf{0.5954} \\ \noalign{\vskip -6pt}
& & AR & 0.7904 & 0.8088 & 0.8366 & 0.8688 & 0.5211 & 0.5196 & 0.5541 & 0.6060 & 0.6544 & 0.6466 & 0.6478 & 0.6718 \\ \noalign{\vskip -1pt}
\Xhline{1.1pt}

\end{tabular}%
}
\vspace{0.1em}

\label{tab:1}
\end{table*}

\begin{table*}[!b]
\centering
\caption{Table for the medium demand traffic; $M=500$, $a=1.6$, $H=0.7$}

\resizebox{\textwidth}{!}{%
\begin{tabular}{!{\vrule width 1.1pt}  M{0.6cm}|M{0.6cm}|M{1cm} !{\vrule width 1.1pt}  *{3}{c|} c !{\vrule width 1.1pt} *{3}{c|} c!{\vrule width 1.1pt} *{3}{c|} c!{\vrule width 1.1pt}   }

\Xhline{1.1pt}
\multicolumn{3}{ !{\vrule width 1.1pt}  c !{\vrule width 1.1pt} }{ Granularity } & \multicolumn{4}{c  !{\vrule width 1.1pt}}{\textbf{10ms}} & \multicolumn{4}{c  !{\vrule width 1.1pt}}{\textbf{100ms}} & \multicolumn{4}{c !{\vrule width 1.1pt} }{\textbf{1000ms (1s)}} \\
\hline
\multicolumn{3}{ !{\vrule width 1.1pt}  c !{\vrule width 1.1pt} }{ Sequence Length } & 64 & 128 & 256 & 512 & 64 & 128 & 256 & 512 & 64 & 128 & 256 & 512 \\
\Xcline{1-4}{0.4pt}
\Xcline{4-15}{1.1pt}

&  & INF & \textbf{0.5514} & 0.5592 & \textbf{0.5439} & \textbf{0.5441} & \textbf{0.7380} & \textbf{0.7370} & \textbf{0.7385} & 0.7433  & \textbf{0.8064} & \textbf{0.8086} & \textbf{0.8062} & 0.8059 \\ \noalign{\vskip -6pt}
& 1 & FAR & 0.5811 & \textbf{0.5581} & 0.5498 & \textbf{0.5441} & 0.7931 & 0.7609 & 0.7467 & \textbf{0.7393} & 0.8677 & 0.8313 & 0.8164 & \textbf{0.8050}  \\ \noalign{\vskip -6pt}
& & AR & 0.6513 & 0.6559 & 0.6613 & 0.6696  & 0.7813 & 0.7639 & 0.7551 & 0.7522 & 0.9661 & 0.8256 & 0.8184 & 0.8105  \\ \noalign{\vskip -1pt}
\Xcline{2-3}{0.4pt}
\Xcline{4-15}{0.4pt}

\multirow{5}{*}{\rotatebox{90}{Prediction Length}}&
& INF & \textbf{0.8708} & 0.8789 & \textbf{0.9027} & 0.9063 &  \textbf{0.9379} & \textbf{0.9406} & \textbf{0.9397} & 0.9386 & \textbf{0.9584} & \textbf{0.9574} & \textbf{0.9558} & \textbf{0.9477}  \\ \noalign{\vskip -6pt}
 & 12 & FAR & 0.8984 & \textbf{0.8728} & 0.9417 & \textbf{0.8596} & 1.0012 & 0.9632 & 0.9490 & \textbf{0.9366}  & 1.0155 & 0.9894 & 0.9706 & 0.9524 \\ \noalign{\vskip -6pt}
& & AR & 0.9376 & 0.9391 & 0.9511 & 0.9633 & 0.9868 & 0.9689 & 0.9632 & 0.9640 & 0.9881 & 0.9787 & 0.9723 & 0.9624 \\ \noalign{\vskip -1pt}
\Xcline{2-3}{0.4pt}
\Xcline{4-15}{0.4pt}

&  & INF & 0.9589 & 0.9362 & \textbf{0.9653} & 0.9622  & \textbf{0.9765} & \textbf{0.9659} & \textbf{0.9681} & 0.9737  & \textbf{0.9781} & \textbf{0.9883} & \textbf{0.9728} & 0.9696 \\ \noalign{\vskip -6pt}
& 24 & FAR & \textbf{0.9454} & \textbf{0.9234} & 1.0020 & \textbf{0.9109} & 1.0327 & 0.9977 & 0.9864 & \textbf{0.9730} & 1.0245 & 1.0048 & 0.9877 & \textbf{0.9692} \\ \noalign{\vskip -6pt}
& & AR & 0.9655 & 0.9622 & 0.9694 & 0.9774 & 1.0250 & 1.0014 & 0.9928 & 0.9905 & 1.0080 & 0.9974 & 0.9878 & 0.9763 \\ \noalign{\vskip -1pt}
\Xcline{2-3}{0.4pt}
\Xcline{4-15}{0.4pt}

&  & INF & 0.9883 & 1.0044 & \textbf{1.0074} & 0.9835 & \textbf{0.9977} & \textbf{0.9977} & \textbf{0.9887} & 1.0107 & \textbf{0.9891} & \textbf{0.9991} & \textbf{0.9884} & \textbf{0.9794} \\ \noalign{\vskip -6pt}
& 48 & FAR & \textbf{0.9777} & \textbf{0.9583} & 1.0383 & \textbf{0.9450} & 1.0630 & 1.0260 & 1.0120 & \textbf{0.9990} & 1.0359 & 1.0166 & 0.9976 & 0.9803 \\ \noalign{\vskip -6pt}
& & AR & 0.9876 & 0.9808 & 0.9846 & 0.9880 & 1.0567 & 1.0259 & 1.0137 & 1.0052 & 1.0270 & 1.0125 & 0.9975 & 0.9840 \\ \noalign{\vskip -1pt}
\Xhline{1.1pt}

\end{tabular}%
}
\vspace{0.1em}
\label{tab:2}
\end{table*}

Demand is initially generated at a granularity of 10 \textit{ms}, and subsequently aggregated to lower granularities of 100 \textit{ms} and 1000 \textit{ms}. The datasets ultimately include 60,000 samples for each level of granularity. We use $70\%$ of the samples for training, $10\%$ for validation, and $20\%$ for testing. As shown in Tables \ref{tab:1}, \ref{tab:2}, and \ref{tab:3}, we consider four different input sequence lengths: 64, 128, 256, and 512, to forecast 1, 12, 24, and 48 steps. We use \textit{ARIMA} method from \textit{statsmodel} library \cite{seabold2010statsmodels} of \textit{Python} and \textit{arfima} package \cite{veenstra2013persistence} of \textit{R} for ARIMA and FARIMA models, respectively. For both ARIMA and FARIMA models, $p$ and $q$ are chosen as 2 and 0 respectively after following the third step given in \ref{arfar}.  The Informer models are trained utilizing the publicly available codebase provided by the original authors \cite{zhou2021informer}, using an NVIDIA RTX A2000 GPU with 12 GB of memory. Subsequently, the forecasting tasks are performed using an NVIDIA RTX 500 Ada GPU with 4 GB of memory. On top of the code base, we additionally implemented ``millisecond" time embeddings and updated feature offsets accordingly. Besides that, the default configuration is used.\\
We follow MSE metric, $MSE=\frac{1}{n}\sum_{i=1}^n(y-\hat{y})^2$, $n$ being the number of test samples. Out of 144 possible cases, the Informer models achieve higher accuracy in 83 cases, FARIMA in 55 cases, and ARIMA in 5 cases. In only one case, Informer and FARIMA yield the same accuracy. It is worth noting that, as the long-range dependency decreases (as $H$ decreases), forecasting the future gets more erroneous.

\section{Conclusion} \label{conc}
In this work, we highlighted the need for demand forecasting to enable traffic-driven beam-hopping in LEO satellite constellations and modeled the forward link using a second-order self-similar traffic with millisecond time granularity. FARIMA was used as a benchmark due to its asymptotic self-similarity properties, yet its reliance on sequence-specific parameter tuning of $p$, $q$, and fitting per input limits practicality for in-orbit use. To address this, we proposed a transformer-based forecasting method with Prob-Sparse self-attention and demonstrated its advantages over FARIMA through extensive simulations under the MSE metric.
\FloatBarrier

\begin{table*}[!t]
\centering
\caption{Table for the low demand traffic; $M=250$, $a=1.9$, $H=0.55$}
\resizebox{\textwidth}{!}{%
\begin{tabular}{!{\vrule width 1.1pt}  M{0.6cm}|M{0.6cm}|M{1cm} !{\vrule width 1.1pt}  *{3}{c|} c !{\vrule width 1.1pt} *{3}{c|} c!{\vrule width 1.1pt} *{3}{c|} c!{\vrule width 1.1pt}   }

\Xhline{1.1pt}
\multicolumn{3}{ !{\vrule width 1.1pt}  c !{\vrule width 1.1pt} }{ Granularity } & \multicolumn{4}{c  !{\vrule width 1.1pt}}{\textbf{10ms}} & \multicolumn{4}{c  !{\vrule width 1.1pt}}{\textbf{100ms}} & \multicolumn{4}{c !{\vrule width 1.1pt} }{\textbf{1000ms (1s)}} \\
\hline
\multicolumn{3}{ !{\vrule width 1.1pt}  c !{\vrule width 1.1pt} }{ Sequence Length } & 64 & 128 & 256 & 512 & 64 & 128 & 256 & 512 & 64 & 128 & 256 & 512 \\
\Xcline{1-4}{0.4pt}
\Xcline{4-15}{1.1pt}

&  & INF & \textbf{0.5432} & 0.5447 & 0.5277 & 0.5309 & \textbf{0.9155} & \textbf{0.8968} & 0.9310 & 0.9253  & 0.9727 & 0.9746 & 0.9630 & 0.9951 \\ \noalign{\vskip -6pt}
& 1 & FAR & 0.5461 & \textbf{0.5268} & \textbf{0.5206} & \textbf{0.5142} & 0.9568 & 0.9138 & \textbf{0.8967} & \textbf{0.8929} & \textbf{0.8677} & \textbf{0.8313} & \textbf{0.8164} & \textbf{0.8050}  \\ \noalign{\vskip -6pt}
& & AR & 0.6186 & 0.6214 & 0.6185 & 0.6195  & 0.9322 & 0.9064 & 0.8970 & 0.8946 & 0.9921 & 0.9709 & 0.9638 & 0.9565  \\ \noalign{\vskip -1pt}
\Xcline{2-3}{0.4pt}
\Xcline{4-15}{0.4pt}

\multirow{5}{*}{\rotatebox{90}{{Prediction Length}}}&
& INF & \textbf{0.8915} & 0.8898 & 0.9148 & 0.9859 &  \textbf{0.9850} & \textbf{0.9961} & \textbf{0.9915} & \textbf{0.9946} & \textbf{1.0098} & 1.0131 & 1.0156 & 1.0225  \\ \noalign{\vskip -6pt}
 & 12 & FAR & 0.8980 & \textbf{0.8755} & \textbf{0.8686} & \textbf{0.8561} & 1.0515 & 1.0193 & 1.0025 & 0.9994  & 1.0183 & \textbf{0.9817} & \textbf{0.9659} & \textbf{0.9550} \\ \noalign{\vskip -6pt}
& & AR & 0.9336 & 0.9364 & 0.9314 & 0.9316 & 1.0270 & 1.0053 & 0.9977 & 0.9986 & 1.0319 & 1.0228 & 1.0221 & 1.0167 \\ \noalign{\vskip -1pt}
\Xcline{2-3}{0.4pt}
\Xcline{4-15}{0.4pt}

&  & INF & \textbf{0.9505} & 0.9502 & 0.9563 & 0.9665  & \textbf{0.9951} & \textbf{0.9979} & \textbf{1.0040} & 1.0170  & \textbf{1.0120} & 1.0322 & \textbf{1.0254} & 1.0252 \\ \noalign{\vskip -6pt}
& 24 & FAR & 0.9638 & \textbf{0.9454} & \textbf{0.9415} & \textbf{0.9289} & 1.0524 & 1.0267 & 1.0110 & 1.0086 & 1.0473 & 1.0335 & 1.0292 & \textbf{1.0184} \\ \noalign{\vskip -6pt}
& & AR & 0.9800 & 0.9785 & 0.9711 & 0.9696 & 1.0386 & 1.0160 & 1.0072 & \textbf{1.0074} & 1.0366 & \textbf{1.0281} & 1.0267 & 1.0205 \\ \noalign{\vskip -1pt}
\Xcline{2-3}{0.4pt}
\Xcline{4-15}{0.4pt}

&  & INF & \textbf{0.9995} & 1.0016 & 1.0028 & 1.0101 & \textbf{1.0017} & \textbf{1.0142} & 1.0219 & 1.0377 & \textbf{1.0172} & \textbf{1.0316} & 1.0349 & 1.0283 \\ \noalign{\vskip -6pt}
& 48 & FAR & 0.9995 & \textbf{0.9824} & \textbf{0.9769} & \textbf{0.9676} & 1.0533 & 1.0288 & 1.0153 & 1.0124 & 1.0498 & 1.0374 & 1.0327 & \textbf{1.0216} \\ \noalign{\vskip -6pt}
& & AR & 1.0079 & 1.0005 & 0.9910 & 0.9890 & 1.0454 & 1.0227 & \textbf{1.0130} & \textbf{1.0120} & 1.0442 & 1.0348 & \textbf{1.0308} & 1.0228 \\ \noalign{\vskip -1pt}
\Xhline{1.1pt}

\end{tabular}%
}
\vspace{0.1em}
\label{tab:3}
\end{table*}

\FloatBarrier

As future work, we plan to conduct sensitivity analysis with time-invariant Hurst values and investigate lighter models than Informer using more resource-constrained hardware. Additionally, we are going to investigate the synchronization process and the relevant time constraints for metadata transfer between two LEO satellites.
\FloatBarrier

\section*{Acknowledgment}
This work was supported in part by MDA; in part by the Consortium de Recherche et d’innovation en Aérospatiale au Québec (CRIAQ); and in part by the Natural Sciences and Engineering Research Council of Canada (NSERC).
\bibliographystyle{IEEEtran} 
\bibliography{main}

\begin{thebibliography}{10}
\providecommand{\url}[1]{#1}
\csname url@samestyle\endcsname
\providecommand{\newblock}{\relax}
\providecommand{\bibinfo}[2]{#2}
\providecommand{\BIBentrySTDinterwordspacing}{\spaceskip=0pt\relax}
\providecommand{\BIBentryALTinterwordstretchfactor}{4}
\providecommand{\BIBentryALTinterwordspacing}{\spaceskip=\fontdimen2\font plus
\BIBentryALTinterwordstretchfactor\fontdimen3\font minus \fontdimen4\font\relax}
\providecommand{\BIBforeignlanguage}[2]{{%
\expandafter\ifx\csname l@#1\endcsname\relax
\typeout{** WARNING: IEEEtran.bst: No hyphenation pattern has been}%
\typeout{** loaded for the language `#1'. Using the pattern for}%
\typeout{** the default language instead.}%
\else
\language=\csname l@#1\endcsname
\fi
#2}}
\providecommand{\BIBdecl}{\relax}
\BIBdecl

\bibitem{yahia2024evolution}
O.~B. Yahia, Z.~Garroussi, O.~Bélanger, B.~Sansò, J.-F. Frigon, S.~Martel, A.~Lesage-Landry, and G.~Karabulut~Kurt, ``Evolution of high-throughput satellite systems: A vision of programmable regenerative payload,'' \emph{IEEE Com. Surveys \& Tutorials}, vol.~27, no.~3, pp. 1565--1597, 2025.

\bibitem{guo2022static}
J.~Guo, L.~Yang, D.~Rincón, S.~Sallent, Q.~Chen, and X.~Liu, ``Static placement and dynamic assignment of {SDN} controllers in {LEO} satellite networks,'' \emph{IEEE Transactions on Network and Service Management}, vol.~19, no.~4, pp. 4975--4988, 2022.

\bibitem{dvbs2x}
{ETSI}, ``{EN 302 307-2 V1.4.1 (2024-08): Digital Video Broadcasting (DVB); Part 2: DVB-S2 Extensions (DVB-S2X)},'' \url{https://www.etsi.org/deliver/etsi_en/302300_302399/30230702/01.04.01_60/en_30230702v010401p.pdf}, Aug. 2024.

\bibitem{chen2024joint}
L.~Chen, L.~Wu, E.~Lagunas, A.~Wang, L.~Lei, S.~Chatzinotas, and B.~Ottersten, ``Joint power allocation and beam scheduling in beam-hopping satellites: A two-stage framework with a probabilistic perspective,'' \emph{IEEE Transactions on Wireless Com.}, vol.~23, no.~10, pp. 14\,685--14\,701, 2024.

\bibitem{lin2022multi}
Z.~Lin, Z.~Ni, L.~Kuang, C.~Jiang, and Z.~Huang, ``Multi-satellite beam hopping based on load balancing and interference avoidance for {NGSO} satellite communication systems,'' \emph{IEEE Transactions on Com.}, vol.~71, no.~1, pp. 282--295, 2023.

\bibitem{wang2024joint}
J.~Wang, C.~Qi, S.~Yu, and S.~Mao, ``Joint beamforming and illumination pattern design for beam-hopping {LEO} satellite communications,'' \emph{IEEE Transactions on Wireless Com.}, vol.~23, no.~12, pp. 18\,940--18\,950, 2024.

\bibitem{al2020traffic}
H.~Al-Hraishawi, E.~Lagunas, and S.~Chatzinotas, ``Traffic simulator for multibeam satellite communication systems,'' in \emph{Advanced Satellite Multimedia Systems Conference \& Signal Processing for Space Com. Workshop (ASMS/SPSC)}, 2020, pp. 1--8.

\bibitem{guo2021gateway}
J.~Guo, D.~Rincon, S.~Sallent, L.~Yang, X.~Chen, and X.~Chen, ``Gateway placement optimization in {LEO} satellite networks based on traffic estimation,'' \emph{IEEE Transactions on Vehicular Technology}, vol.~70, no.~4, pp. 3860--3876, 2021.

\bibitem{bie2019combined}
Y.~Bie, L.~Wang, Y.~Tian, and Z.~Hu, ``A combined forecasting model for satellite network self-similar traffic,'' \emph{IEEE Access}, vol.~7, pp. 152\,004--152\,013, 2019.

\bibitem{leland2002self}
W.~Leland, M.~Taqqu, W.~Willinger, and D.~Wilson, ``On the self-similar nature of {E}thernet traffic (extended version),'' \emph{IEEE/ACM Transactions on Networking}, vol.~2, no.~1, pp. 1--15, 1994.

\bibitem{iturm2370}
{ITU-R}, ``{IMT traffic estimates for the years 2020 to 2030},'' International Telecommunication Union - Radiocommunication Sector (ITU-R), Report M.2370-0, Jul. 2015, available: \url{https://www.itu.int/dms_pub/itu-r/opb/rep/R-REP-M.2370-2015-PDF-E.pdf}.

\bibitem{taqqu1997proof}
M.~S. Taqqu, W.~Willinger, and R.~Sherman, ``Proof of a fundamental result in self-similar traffic modeling,'' \emph{SIGCOMM Comput. Commun. Rev.}, vol.~27, no.~2, p. 5–23, Apr. 1997.

\bibitem{beran2013long}
J.~Beran, Y.~Feng, S.~Ghosh, and R.~Kulik, \emph{Long-Memory Processes: Probabilistic Properties and Statistical Methods}, ser. SpringerLink : B{\"u}cher.\hskip 1em plus 0.5em minus 0.4em\relax Springer Berlin Heidelberg, 2013.

\bibitem{vaswani2017attention}
A.~Vaswani, N.~Shazeer, N.~Parmar, J.~Uszkoreit, L.~Jones, A.~N. Gomez, L.~Kaiser, and I.~Polosukhin, ``Attention is all you need,'' in \emph{International Conference on Neural Information Processing Systems}, 2017, p. 6000–6010.

\bibitem{fournier2023practical}
Q.~Fournier, G.~M. Caron, and D.~Aloise, ``A practical survey on faster and lighter transformers,'' \emph{ACM Comput. Surv.}, vol.~55, no. 14s, Jul. 2023.

\bibitem{zhou2021informer}
H.~Zhou, S.~Zhang, J.~Peng, S.~Zhang, J.~Li, H.~Xiong, and W.~Zhang, ``Informer: Beyond efficient transformer for long sequence time-series forecasting,'' in \emph{Proceedings of the AAAI conference on artificial intelligence}, vol.~35, no.~12, 2021, pp. 11\,106--11\,115.

\bibitem{montanari1997fractionally}
A.~Montanari, R.~Rosso, and M.~S. Taqqu, ``Fractionally differenced {ARIMA} models applied to hydrologic time series: Identification, estimation, and simulation,'' \emph{Water Resources Research}, vol.~33, no.~5, pp. 1035--1044, 1997.

\bibitem{veenstra2013persistence}
J.~Q. Veenstra, \emph{Persistence and Anti-Persistence: Theory and Software}.\hskip 1em plus 0.5em minus 0.4em\relax Ph.D dissertation, The University of Western Ontario (Canada), 2013.

\bibitem{seabold2010statsmodels}
S.~Seabold, J.~Perktold \emph{et~al.}, ``Statsmodels: econometric and statistical modeling with python.'' \emph{SciPy}, vol.~7, no.~1, pp. 92--96, 2010.

\end{thebibliography}

\end{document}